\documentclass[aip, preprint]{revtex4-1}

\usepackage[]{natbib}
\makeatletter \renewcommand\@biblabel[1]{$^{#1}$} \makeatother
\usepackage{hyperref}
\hypersetup{ colorlinks,
    citecolor=blue,
    filecolor=blue,
    linkcolor=blue,
    urlcolor=blue
}

\usepackage[toc,page]{appendix}

\usepackage{graphicx}% Include figure files
\usepackage{dcolumn}% Align table columns on decimal point
\usepackage{bm}% bold math

\usepackage[mathlines]{lineno}% Enable numbering of text and display math
\usepackage{xcolor}
\definecolor{gray}{rgb}{0.6,0.6,0.6}
\definecolor{red}{rgb}{0.85,0,0}
\definecolor{green}{rgb}{0,0.85,0}
\definecolor{blue}{rgb}{0,0,0.85}
\definecolor{beige}{rgb}{0.92,0.87,0.78}
\usepackage[all]{hypcap}    %causes link to figures to go to figure, not caption
\usepackage[utf8]{inputenc}
\usepackage{gensymb} % Degree
\usepackage{booktabs} % Tables
\usepackage{subcaption} % Subfigures
\usepackage{adjustbox}
\usepackage{mathtools, nccmath}
\usepackage{xparse}
\usepackage{algorithm}
\usepackage{algpseudocode}
\usepackage{multirow}
\usepackage{setspace}

\begin{document}
%\linespread{1.25}
\title[Beamlet-free treatment planning]{Beamlet-free optimization for Monte Carlo based treatment planning in proton therapy}
% SUGGESTION OF ALTERNATIVE TITLES
% Beamlet-free optimization for treatment planning: feasibility and first results in proton therapy
% Feasibility study of beamlet-free  proton therapy treatment planning optimization
% Beamlet-free treatment planning (for proton therapy)
% Validation of a beamlet-free optimization algorithm (for treatment planning in proton therapy)

\author{Danah Pross}
\affiliation{ 
Université catholique de Louvain, Institut de Recherche Expérimentale et Clinique (IREC), Center of Molecular Imaging, Radiotherapy and Oncology, Louvain-La-Neuve, 1348, Belgium
}%

\author{Sophie Wuyckens}
\affiliation{ 
Université catholique de Louvain, Institut de Recherche Expérimentale et Clinique (IREC), Center of Molecular Imaging, Radiotherapy and Oncology, Louvain-La-Neuve, 1348, Belgium
}%

\author{Sylvain Deffet}
\affiliation{ 
Université catholique de Louvain, Institut de Recherche Expérimentale et Clinique (IREC), Center of Molecular Imaging, Radiotherapy and Oncology, Louvain-La-Neuve, 1348, Belgium
}%

\author{Edmond Sterpin}
\affiliation{ 
Université catholique de Louvain, Institut de Recherche Expérimentale et Clinique (IREC), Center of Molecular Imaging, Radiotherapy and Oncology, Louvain-La-Neuve, 1348, Belgium
}%
\affiliation{
KU Leuven, Department of Oncology, Laboratory of Experimental Radiotherapy, Leuven, 3000, Belgium
}%
\affiliation{
Particle Therapy Interuniversity Center Leuven - PARTICLE, Leuven, 3000, Belgium
}%

\author{John A.~Lee}
\affiliation{ 
Université catholique de Louvain, Institut de Recherche Expérimentale et Clinique (IREC), Center of Molecular Imaging, Radiotherapy and Oncology, Louvain-La-Neuve, 1348, Belgium
}%

\author{Kevin Souris}
\affiliation{ 
Université catholique de Louvain, Institut de Recherche Expérimentale et Clinique (IREC), Center of Molecular Imaging, Radiotherapy and Oncology, Louvain-La-Neuve, 1348, Belgium
}%
\affiliation{
Ion Beam Applications SA, Louvain-La-Neuve, 1348, Belgium
}

\date{\today}
\begin{abstract}
\textbf{Background:} Dose calculation and optimization algorithms in proton therapy treatment planning often have high computational requirements regarding time and memory. This can hinder the implementation of efficient workflows in clinics and prevent the use of new, elaborate treatment techniques aiming to improve clinical outcomes like robust optimization, arc and adaptive proton therapy. \\
\textbf{Purpose:} A new method, namely, the beamlet-free algorithm, is presented to address the aforementioned issue by combining Monte Carlo dose calculation and optimization into a single algorithm and omitting the calculation of the time-consuming and costly dose influence matrix.\\
\textbf{Methods:} The beamlet-free algorithm simulates the dose in proton batches of randomly chosen spots and evaluates their relative impact on the objective function at each iteration. Based on the approximated gradient, the spot weights are then updated and used to generate a new spot probability distribution. The beamlet-free method is compared against a conventional, beamlet-based treatment planning algorithm on a brain case.\\
\textbf{Results:} The beamlet-free algorithm maintained a comparable plan quality while reducing the computation time by 70\% and the peak memory usage by 95\%.\\
\textbf{Conclusion:} The implementation of a beamlet-free treatment planning algorithm for proton therapy is feasible and capable of achieving a considerable reduction of time and memory requirements.\\

\end{abstract}


\maketitle
\section{Introduction}
In proton therapy, the state-of-the-art is to treat patients with intensity modulated proton therapy\cite{Kooy_2015} (IMPT) using pencil beam scanning. Treatment planning typically involves the precalculation of individual beamlet dose distributions and subsequent optimization of their respective intensity. The total dose to voxel $i$ is then given as the linear combination of these elementary dose distributions
\begin{linenomath*}
\begin{equation}
    d_{i}=\sum_j D_{ij} x_j 
\end{equation}
\end{linenomath*}
with $D_{ij}$ as the individual beamlet dose contribution of spot $j$ to voxel $i$ and $x_j$ its corresponding intensity (`spot weight'). The dose influence matrix $D$ is the concatenation of the beamlet dose distributions and usually reaches considerable size, due to the large number of spots in IMPT.\\
While dose calculations can be performed using fast analytical dose calculation algorithms \cite{Hong_1996}, their accuracy falls short of the more elaborate Monte Carlo methods \cite{Paganetti_2012,Taylor2017}. The drawback of MC dose engines is their high demand of computational resources that, especially for large numbers of spots, can prolong treatment planning time considerably. It is therefore an ongoing effort to reduce computation time for MC methods and several fast MC dose engines have been introduced, often based on CPU (Central Processing Unit) or GPU (Graphical Processing Unit) parallelization \cite{Fippel2004,Tourovsky_2005, Kohno_2011,Jabbari2014,Souris2016,Schiavi_2017}.\\
The storage of the beamlet dose distributions in the dose influence matrix, necessary for optimization, also puts considerable strain on the the computers memory capacities \cite{Ziegenhein_2008}.\\
Besides the time expended on the repeated calls to the dose calculation engine, the optimization algorithm also places high demands on the computational resources and prolongs computation time, due to the very costly multiplications with the large dose influence matrix involved in the gradient computation.\\
These high computational demands, both with respect to computation time and memory usage, are hindrances in clinical routines that require to be addressed to enable efficient workflows when a MC dose engine is involved. These problems become additionally exacerbated in study cases that require a large number of beamlets, such as robust optimization, adaptive proton therapy or arc proton therapy. These increasingly elaborate methods seek to improve treatment quality by reducing dose to organs at risk (OAR) and ensuring target coverage, but are limited by the current computational demands.\\
The aim of this work is to present a new method of treatment plan optimization that substantially reduces the required computation time and memory, while maintaining the plan quality. The approach combines MC dose calculation and spot weight optimization into a single efficient algorithm. This algorithm is referred to as beamlet-free algorithm  since it does not require the computation of individual beamlet dose distributions. The method is applied here to proton therapy with pencil beam scanning, where the number of beamlets is typically higher than in photon therapy and where the superior accuracy of MC methods over analytical dose calculation algorithms is especially beneficial due to the steep dose gradients involved.\\ 
\section{Methods}
\subsection{Cost function}
Following a classical approach, the objectives penalize deviations of the dose from a reference value in a region of interest (ROI)\cite{Oelfke2001}. For example, to enforce a prescribed minimum dose $b_{\min}$, the following objective term may be minimized:
\begin{linenomath*}
\begin{equation}
    f_{\min}(\bm{d}) = \frac{1}{N_r} \sum_{i=1}^{N_r} \min \left( 0, d_i(\bm{x}) - b_{\min}\right)^2 \enspace ,
    \label{eq:fmin}
\end{equation}
\end{linenomath*}
where $N_r = |ROI|$ is the number of voxel in the ROI.
A similar objective may be used to enforce a maximum dose $b_{max}$:
\begin{linenomath*}
\begin{equation}
    f_{\max}(\bm{d}) = \frac{1}{N_r} \sum_{i=1}^{N_r} \max \left( 0, d_i(\bm{x}) - b_{\max} \right)^2 \enspace .
    \label{eq:fmax}
\end{equation}
\end{linenomath*}
Several objectives may be defined. Hence, the general treatment plan optimization problem takes the following form:
\begin{linenomath*}
\begin{eqnarray}
\min && F(\bm{x}) = \sum_k f_k(\bm{d})\label{eq:opti_problem}\\
 & \mathrm{s.t. } & \bm{x} \geq 0 \enspace , \nonumber\\
\end{eqnarray}
\end{linenomath*}
where the index $k$ iterates over the objectives and $f_k(\bm{d})$ are either $f_{\min}(\bm{d})$ or $f_{\max}(\bm{d})$ depending on the objective type.

\subsection{Beamlet-free algorithm}
The idea of beamlet-free optimization is based on continually and incrementally evaluating the objective function while the dose is being simulated. Instead of computing each individual beamlet dose distribution, the dose is continually scored during the MC simulation of the proton batches and directly added to a total dose distribution. Consequently, spot weights in their original sense as modulation factors cannot exist anymore. The equivalent of the original spot weights with respect to the final dose output would be the number of protons simulated towards each spot. To optimize this quantity during the ongoing dose simulation, a set of unnormalized probabilities $p$ is introduced, that are used to generate a spot probability distribution $P$.\\
In each iteration of the proposed algorithm, a spot is randomly sampled from this probability distribution $P$. A batch of protons is simulated towards the chosen spot and the impact on the objective function is calculated to estimate the gradient of the objective function with respect to the probability of the chosen spot.\\
After all protons of the batch have been simulated, the estimated gradient, with respect to the sampled spot, is used to update the spot probability, similarly to Stochastic Gradient Descent (SGD) \cite{Robbins1951}.\\
This yields Algorithm \ref{algo:BLFree}, where $p_j$ is the probability for a spot $j$, $P$ is the corresponding probability distribution, $h$ is a step size, $c$ is the batch cost calculated as 
\begin{linenomath*}
\begin{equation}
c = \sum_{i}\sum_{k} F(d_{i,k-1})-F(d_{i,k})\label{eq:cost_update}
\end{equation}
\end{linenomath*}
with $i$ an index iterating over all voxel, $k$ an index over all proton contributions in a batch, $F(\cdot)$ the objective function as described in equation \ref{eq:opti_problem} and $d_{i,k-1}$ and $d_{i,k}$ as the normalized dose to voxel $i$ prior to and after the contribution of particle $k$. The particle cost calculated in \ref{eq:cost_update} corresponds to the finite difference approximation of the objective function gradient.\\
The normalization of the dose during the cost computation is necessary to evaluate the relative impact of simulating particles to a given spot $j$. The total, unnormalized dose is multiplied with a normalization coefficient $n(\bm{d})=\frac{\hat{d}}{b}$ where $\hat{d}$ is the mean target dose and $b$ is the prescribed target dose.\\
\begin{algorithm}[H]
    \caption{Beamlet-free algorithm}
    \label{algo:BLFree}
    \begin{algorithmic}[1]
            \State Initialize spot probability vector $\bm{p}$ to unit values
            \State Generate probability distribution $P(\bm{p})$
            \State Set $n_{\text{simulated}} \leftarrow 0$
            \While{$n_{\text{simulated}} \le n_{\text{total}}$}
                \State  Sample spot $j$ from $P$
                \State Initiate a batch of $n_{\text{batch}}$ particles for spot $j$
                \State Initialize cost $c \leftarrow 0$ 
                \For{$k \le n_{\text{batch}}$}
                    \State Simulate particle $k$
                    \State Update cost $c \leftarrow c + \sum_{i,k}F(d_{i,k-1})-F(d_{i,k})$
                    \State $n_{\text{simulated}} \leftarrow n_{\text{simulated}} + 1$
                \EndFor
                \State Update spot probability: $p_j\leftarrow p_j-hc$
            \EndWhile
    \end{algorithmic}
\end{algorithm}
The quantity to be optimized in this procedure is the spot probability $\bm{p}$, which is stochastically linked to the number of protons simulated toward the corresponding spots, the spot weight $\bm{x}$. The objective function is evaluated over the total dose distribution $\bm{d}$, which depends on $\bm{x}$, and only indirectly on $\bm{p}$. The finite difference approximation in \ref{eq:cost_update} concerns the gradient of the objective function with respect to $\bm{x}$.\\
There are therefore two separate stochastic processes taking place, the MC simulation generating the dose distribution and the spot sampling at the beginning of each iteration which causes the distribution of protons per spot, $\bm{x}$, and the probability distribution $P$ to converge towards the same final distribution.\\
The beamlet-free algorithm was implemented in C around the existing, fast Monte Carlo code MCsquare \cite{Souris2016}.\\

\subsection{Phantom data}
\begin{figure*}[t]
    \begin{center}
        \includegraphics[height=8.5cm]{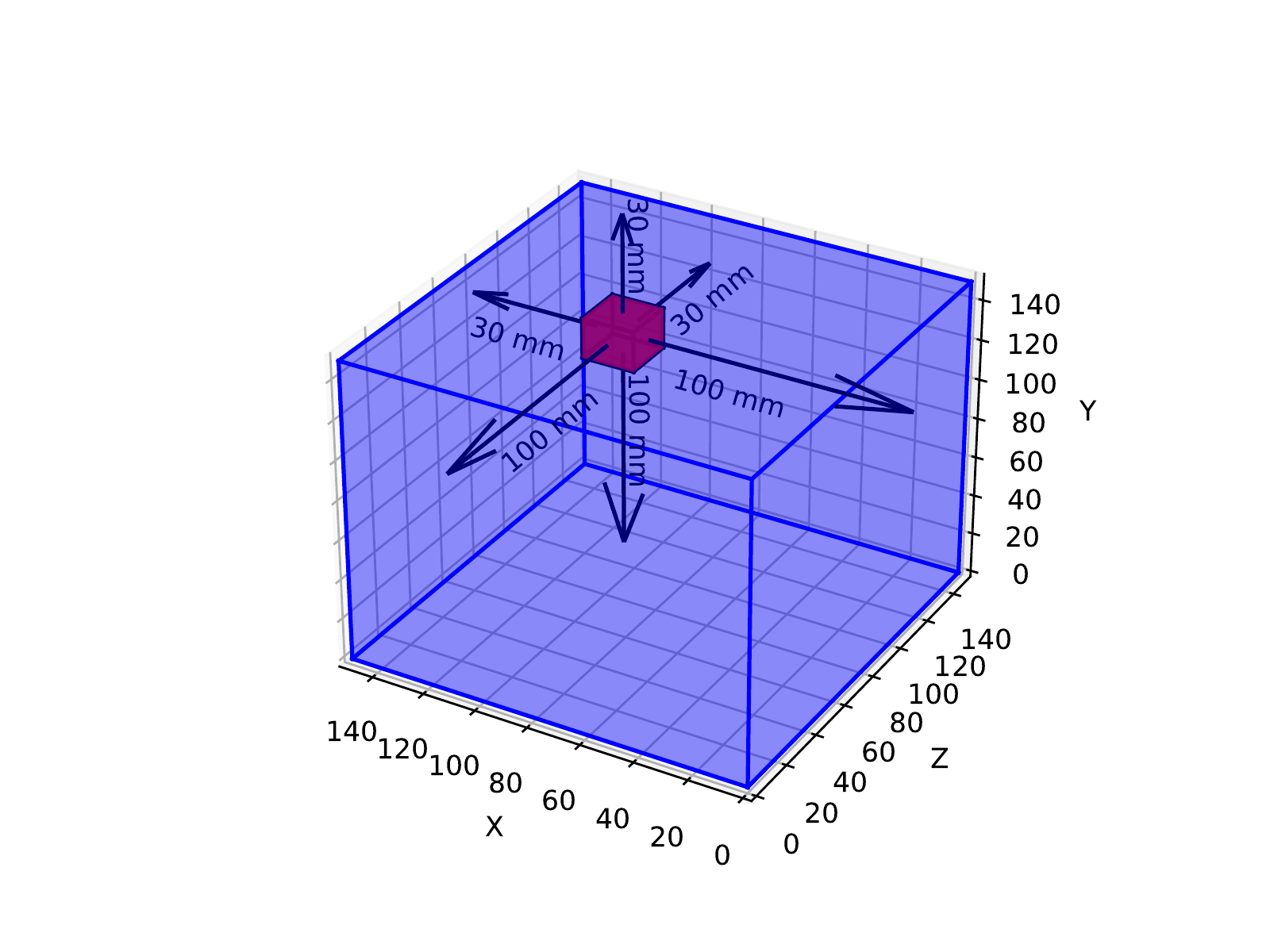}
    \caption{Water phantom used for the scaling analysis. An examplary target region with side length of 20 mm is represented in red.}
    \label{fig:Phantom}
    \end{center}
\end{figure*}

The beamlet-free method is expected to improve both computation time and memory usage when large numbers of beamlets are involved. To highlight its benefits and limitations, a complexity analysis was performed. Simulations are performed on a water phantom containing a target region of various sizes. The phantom is represented in figure \ref{fig:Phantom}. The multiple target sizes result in geometries with varying numbers of spots. Spot and layer spacing are set to 4 mm and a target margin of 6 mm is applied. The influence of the number of spots on computation time and memory usage are investigated.\\
A single beam with couch and gantry angles of $0\degree$ is used for the analysis. The water phantom has dimensions of 150x150x150 mm$^3$ and the target region is cubic, with the center located at a depth of 110 mm in beam direction and side lengths ranging between 20 mm and 60 mm. A representation of the water phantom can be found in figure \ref{fig:Phantom}.\\
Simulations are repeated 5 times to estimate the impact of statistical fluctuations.
\subsection{Patient data}
To demonstrate the high potential of the beamlet-free algorithm, we chose to benchmark it against a conventional, beamlet-based gradient descent optimization algorithm. A brain case (base-of-skull tumor) was chosen for the comparison. Local ethics committee approval was obtained for the use of the patient data. A treatment plan was generated using a three-beam combination at a fixed gantry angle G $= 90^{\circ}$ and rotating the couch angle C at $0^{\circ}$, $270^{\circ}$ and $315^{\circ}$. A total of 9812 spots compose the plan. The ROIs upon which optimization objectives are placed are the clinical target volume (CTV), optical chiasm (OC), brain stem (BS), right optical nerve (RON) and left optical nerve (LON). The optimization objectives are listed in table \ref{Tab:Objectives}.\\
\begin{table}[h!]
\begin{center}
\caption{Optimization objectives}
\label{Tab:Objectives}
\begin{tabular}{ c | c | c | c}
 ROI & Objective Type & Dose limit & Weight\\ 
 \hline
 PTV & $D_{\text{min}}$ & 54.0 & 5\\  
 PTV & $D_{\text{max}}$ & 54.0 & 5\\    
 OC & $D_{\text{max}}$ & 60.0 & 1\\
 BS & $D_{\text{max}}$ & 55.0 & 1\\
 RON & $D_{\text{max}}$ & 55.0 & 1\\
 LON & $D_{\text{max}}$ & 55.0 & 1
\end{tabular}
\end{center}
\end{table}
The treatment plan optimization is performed using the in-house developed, research treatment planning system OpenTPS\cite{openTPS2023} which can incorporate both the beamlet-free method as well as established optimization algorithms such as second order methods like the BFGS algorithm \cite{Broyden1970, Shanno1970, Goldfarb1970, Fletcher1970}.
\subsection{Optimization parameters}
The total number of particles needed in the Monte Carlo simulation, depend on the number of spots. For the water phantom, a number of 40 million primary protons was chosen, which was shown to result in good statistical accuracy for all cases. For the patient case, which was both larger and more complex, a higher number of 200 million particles was chosen. All other parameters were chosen identically for the different simulations.\\
The beamlet-free algorithm simulates the primary protons in batches of 16 corresponding to $12.5\cdot 10^6$ iterations of the first order algorithm for the patient case. For the comparison, a beamlet matrix with $5\cdot 10^4$ protons per beamlet is simulated and subsequently optimized using the scipy \cite{scipy} FORTRAN implementation (wrapped in Python) of the second-order optimization algorithm L-BFGS \cite{Liu1989} (Limited-memory Broyden-Fletcher-Goldfarb-Shanno algorithm). The optimizer continues until it meets the tolerance criterion which is defined as the absolute difference in ojective function values between iterations and set to $10^{-4}$. A final dose computation with 40 million protons is performed to match the statistical accuracy of the beamlet-free method. Both the beamlet calculation and the final dose computation are performed using the MCsquare dose engine that is also utilized in the beamlet-free algorithm. The number of primaries for the MC dose calculation in the beamlet-free algorithm was intentionally picked so that the optimized plan quality would reach the one obtained with the beamlet-based method and consequently ease the comparison.\\
Simulations were performed on a computation server with a total of 80 CPU cores and 526 GB RAM.\\
\section{Results}
\label{sec:results}
\subsection{Scaling analysis}
\begin{figure*}[t]
    \begin{center}
    \begin{subfigure}[t]{0.49\textwidth}
        \centering
        \includegraphics[height=6.5cm]{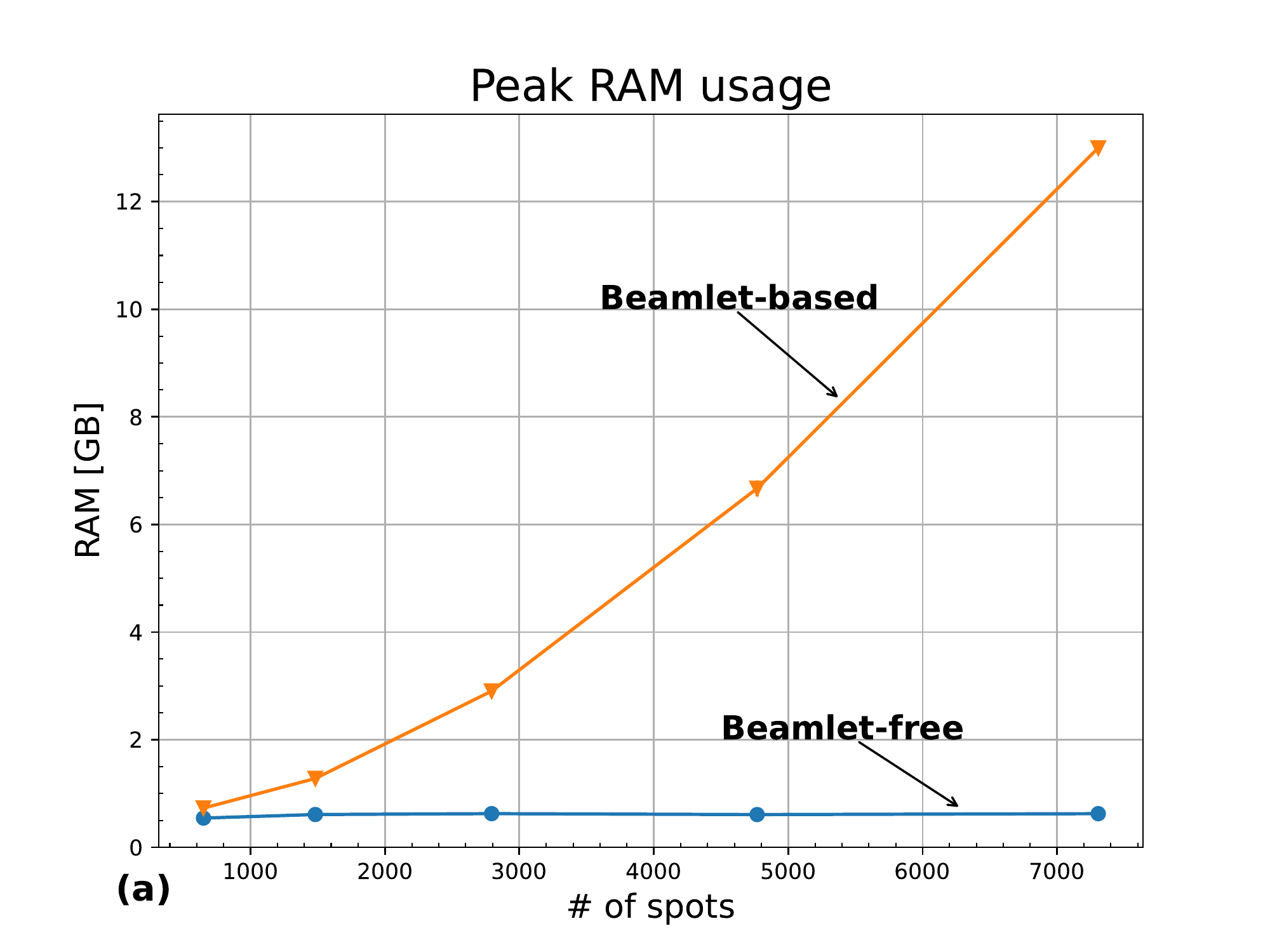}
        \label{fig:memory}
    \end{subfigure}
    \hfill
    \begin{subfigure}[t]{0.49\textwidth}
        \centering
        \includegraphics[height=6.5cm]{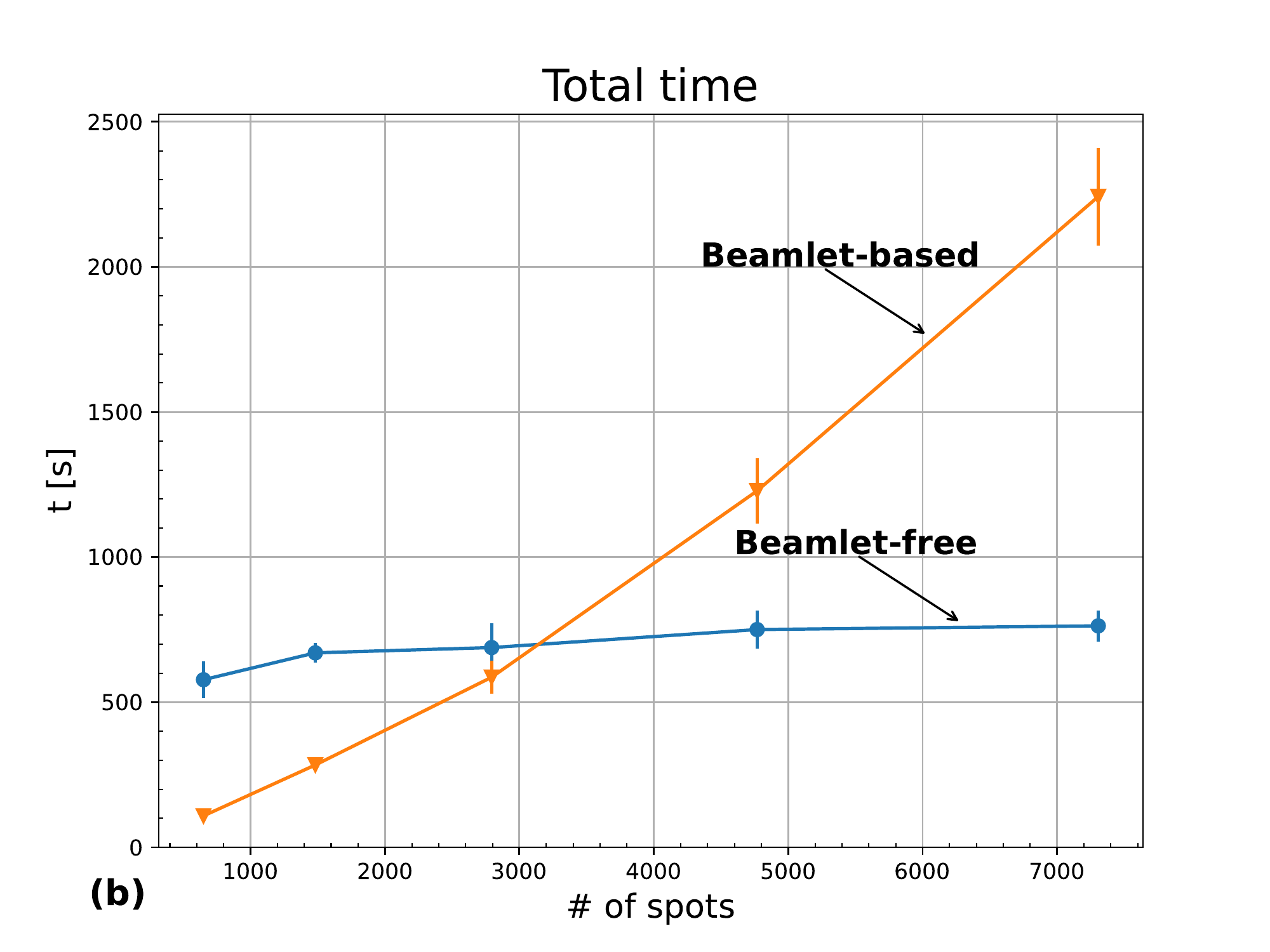}
        \label{fig:time}
    \end{subfigure}
    \caption{Memory usage (a) and computation time (b) required by the beamlet-based and beamlet-free methods versus the number of spots for a water phantom with different target sizes. }
    \label{fig:scalingAnalysis}
    \end{center}
\end{figure*}

Computation time and peak memory usage are plotted against the number of spots for both methods and compared in Figure \ref{fig:scalingAnalysis}. The values are calculated as the mean over 5 runs and the errorbar represents the standard deviation. The difference in target coverage was less than 0.20 Gy in all cases.
\subsection{Patient data}
\begin{figure*}[t]
    \begin{center}
    \begin{subfigure}[t]{0.4\textwidth}
        \centering
        \includegraphics[height=9cm]{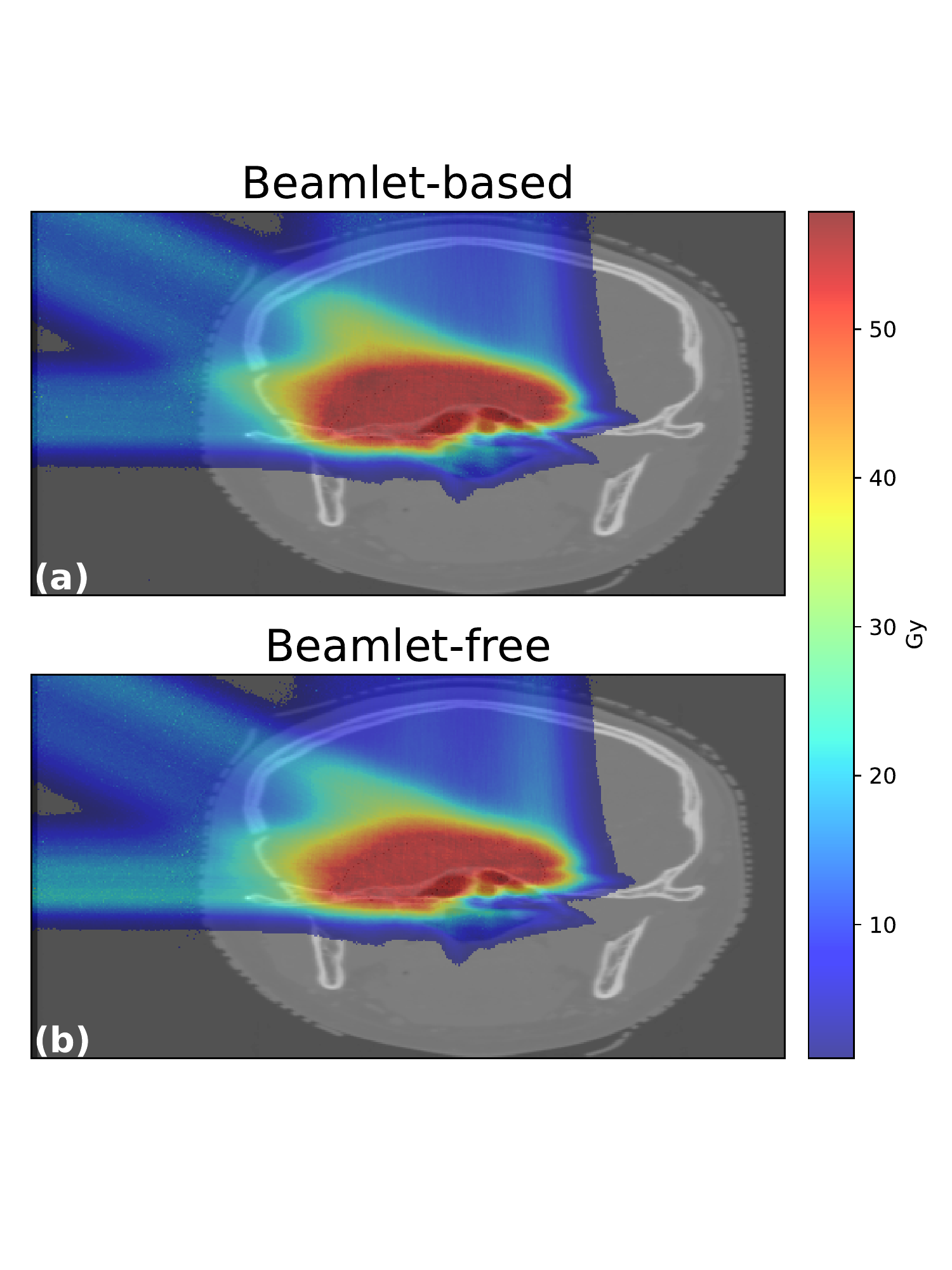}
        \label{fig:DoseDist}
    \end{subfigure}
    \hfill
    \begin{subfigure}[t]{0.59\textwidth}
        \centering
        \includegraphics[height=8.5cm]{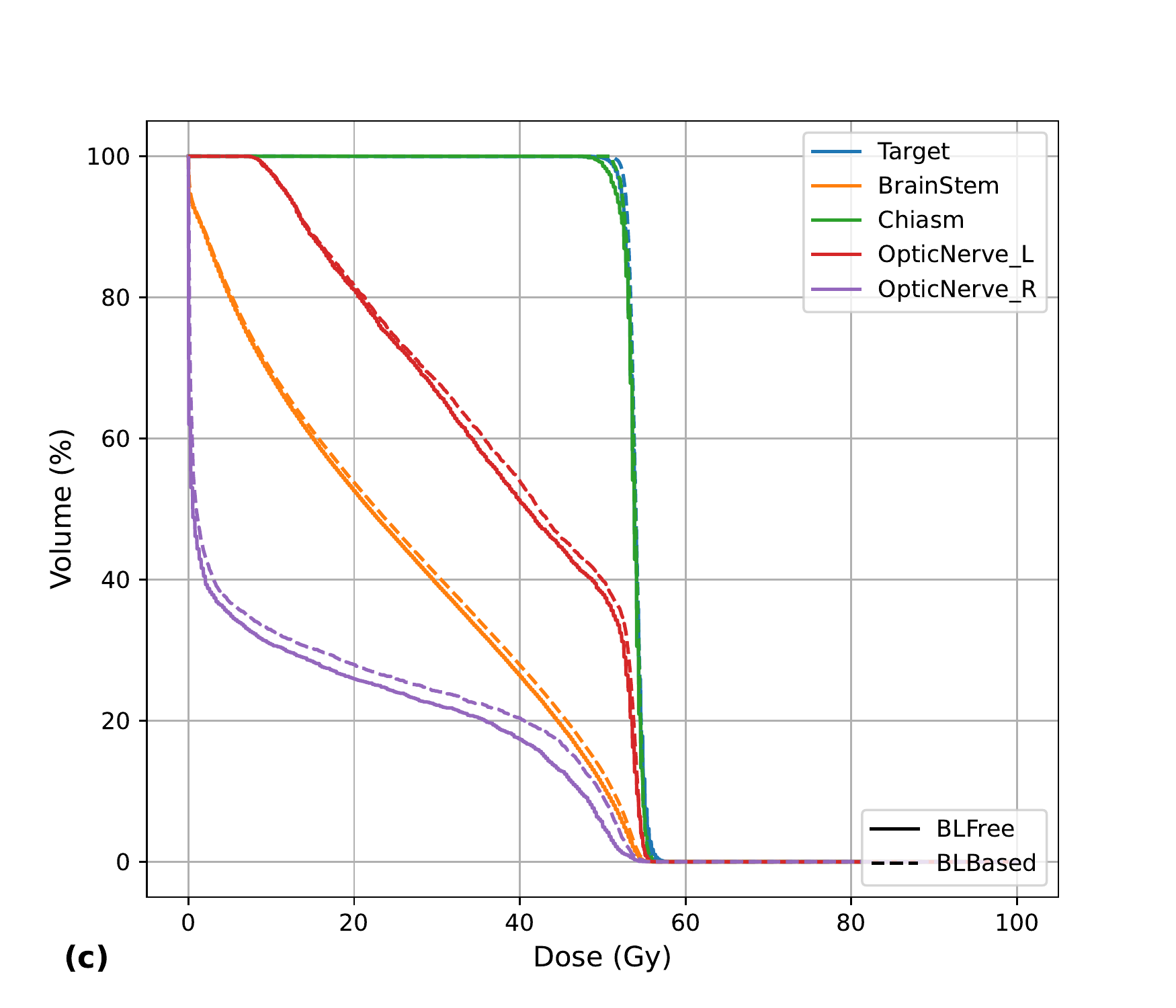}
        \label{fig:DVH}
    \end{subfigure}
    \caption{Comparison of the optimized dose distribution obtained by the beamlet-based (a) and beamlet-free method (b) and the computed dose volume histogram (c) for beamlet-based (dashed) and beamlet-free (solid).}
    \label{fig:DoseComparison}
    \end{center}
\end{figure*}

Dose distributions obtained by both methods are depicted in figure \ref{fig:DoseComparison}. In both cases all OAR objectives could be met and the target D95/D5 is within $5\%$ of the prescription, as it was reported in table \ref{Tab:ClinicalObjectives}.\\
For the OAR, $D_{\text{max}}$ objectives are considered as met, if the $D_{5}$ value of the ROI is below the dose limit. Similarly, $D_{\text{mean}}$ objectives are considered as met, if the $D_{\text{mean}}$ value of the ROI is below the dose limit. The target objectives are considered as met if the $D_{95}$/$D_5$ value is within $\pm 5\%$ of the prescription dose.\\

\begin{figure*}[t]
    \centering
    \begin{subfigure}[t]{0.49\textwidth}
        \centering
        \includegraphics[height=6.5cm]{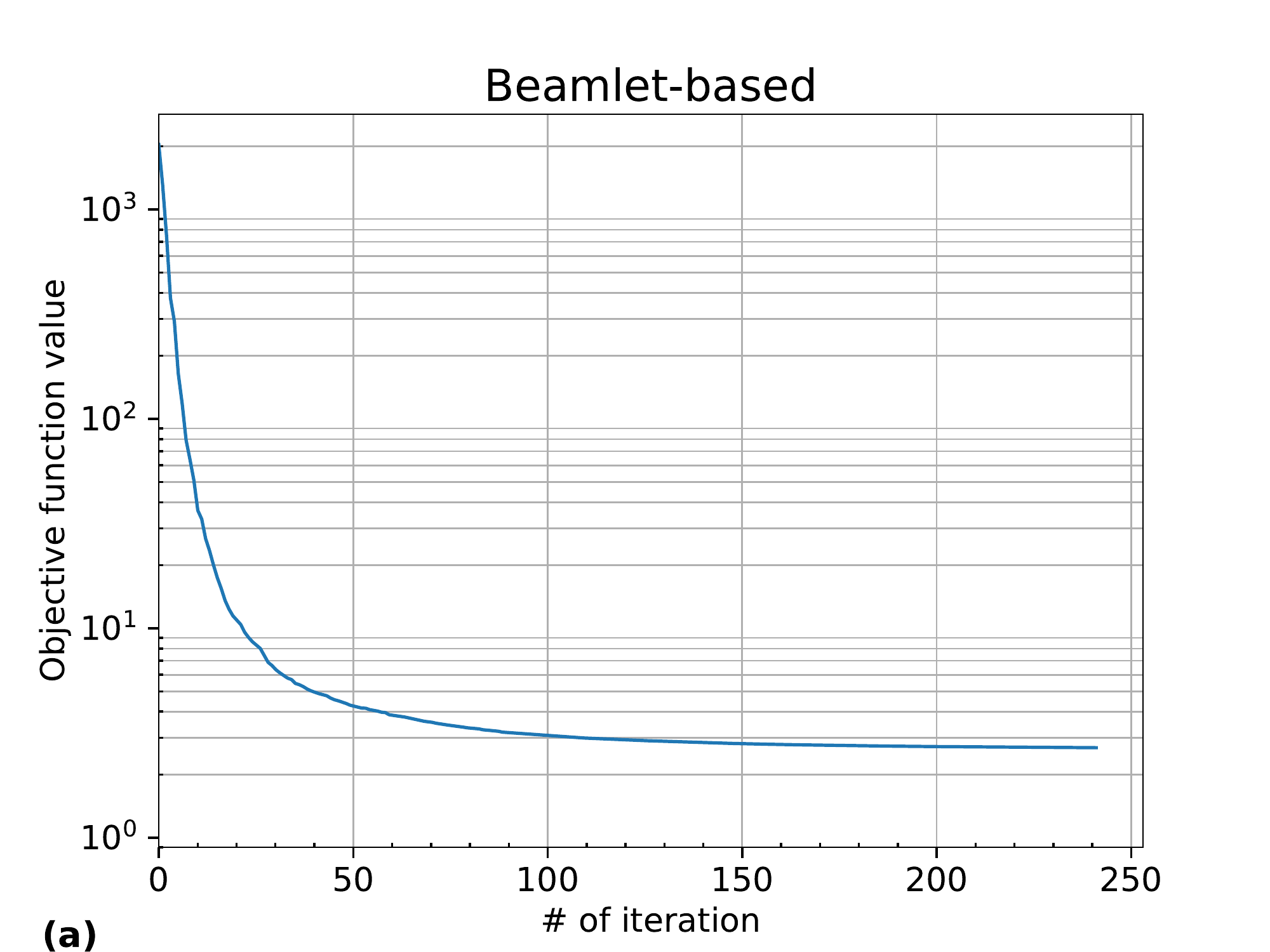}
        \label{fig:ConvergenceBLBased}
    \end{subfigure}
    \hfill
    \begin{subfigure}[t]{0.49\textwidth}
        \centering
        \includegraphics[height=6.5cm]{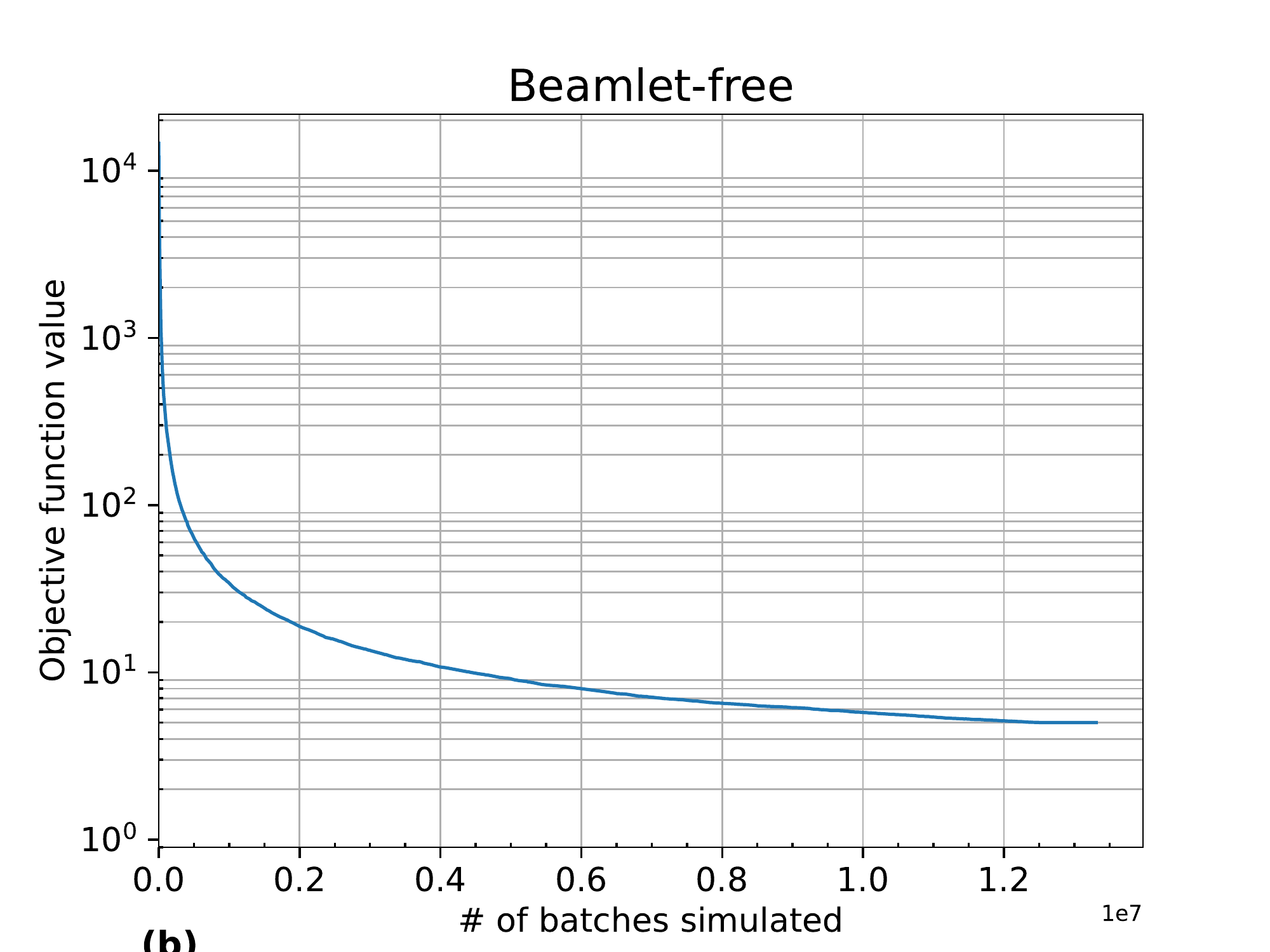}
        \label{fig:ConvergenceBLFree}
    \end{subfigure}
    \caption{(a) Convergence plot of the beamlet-based algorithm (scipy L-BFGS) and (b) convergence plot of the beamlet-free algorithm. }
    \label{fig:ConvergencePlots}
\end{figure*}

%\begin{spacing}{\singlespace}
\begin{table}[htbp]
\begin{center}
\caption{Clinical goals evaluation for the beamlet-based and beamlet-free method.}
\label{Tab:ClinicalObjectives}
\begin{tabular}{ c | c | c | c | c}
 ROI & Criterium & \begin{tabular}{@{}c@{}}Dose \\ limit \\ (Gy)\end{tabular} & \begin{tabular}{@{}c@{}}$D_{\text{Criterium}}$ (Gy)\\ Beamlet-\\based\end{tabular} &  \begin{tabular}{@{}c@{}}$D_{\text{Criterium}}$ (Gy)\\ Beamlet-\\free\end{tabular}\\ 
 \hline
 PTV & $D_{95}$ & 51.3 & 52.73 & 52.28\\  
 PTV & $D_{5}$ & 56.7 & 55.24 & 55.36\\    
 PTV & $D_{98}$ & 52.98 & 52.35 & 51.53\\  
 PTV & $D_{2}$ & 55.08 & 55.58 & 55.87\\    
 OC & $D_{5}$ & 60.0 & 55.08 & 55.09\\
 OC & $D_{2}$ & 60.0 & 55.43 & 55.36\\
 OC & $D_{mean}$ & 50.0 & 53.85 & 53.82\\
 BS & $D_{5}$ & 55.0 & 53.11 & 52.55\\
 BS & $D_{2}$ & 55.0 & 54.03 & 53.82\\
 BS & $D_{mean}$ & 8.50 & 24.79 & 24.16\\
 RON & $D_{5}$ & 55.0 & 51.74 & 49.99\\
 RON & $D_{2}$ & 55.0 & 53.10 & 51.79\\
 LON & $D_{5}$ & 55.0 & 54.57 & 54.58\\
 LON & $D_{2}$ & 55.0 & 54.96 & 55.09
\end{tabular}
\end{center}
\end{table}
%\end{spacing}
Due to the stochastic nature of both methods, small fluctuations in the final dose distributions are to be expected. The particle number and the total number of iterations have to be chosen sufficiently high to ensure an adequate level of confidence. To this end, both methods were run 5 times. The fluctuations on the target coverage,$D_{5}-D_{95}$, were less than $4\%$ for both methods.\\
The objective function convergence over the course of the optimization is shown in figure \ref{fig:ConvergencePlots}. The objective function values are plotted against the number of iteration of the respective optimization algorithm. The relative behaviour of the objective function can be compared, and a similar behaviour of rapidly decreasing objective function values can be observed.\\
The peak memory required during the dose calculation and optimization, could be reduced from $29$ GB for the beamlet-based method to about $1.561$ GB for the beamlet-free method. It is worth noticing that the peak memory for the beamlet-based method was recorded during the import of the individual beamlet dose distributions, compressed as sparce matrices, into the dose influence matrix (stacked as columns). Other steps in the treatment planning process, however, required higher peak memory usage than the beamlet-free method needed during dose calculation and optimization. This peak memory was recorded as about $3.569$ GB. It depends on the used treatment planning system and might be improved independently from the presented algorithm. A memory comparison is drawn in Table \ref{tab:memory}. Besides the peak RAM usage during simulation, it also lists the required storage disk space of the files needed for the dose calculation and optimization. The difference between the two methods with respect to utilized disk space can be explained through the storage of the large dose influence matrix that requires 4.4 GB alone.\\
\begin{table}[htbp]
    \begin{center}
    \caption{Comparison of memory usage for the beamlet-based and beamlet-free method.}
    \label{tab:memory}
    \begin{tabular}{c|c|c}
         & Beamlet-based & Beamlet-free \\
         \hline
         Peak RAM usage [GB] & $29$ & $1.561$\\
         Storage [GB] & $6.4$ & $2.2$\\
    \end{tabular}
    \end{center}
\end{table}\\
The total computation times averaged over 5 runs are reported in table \ref{Tab:CompTime}.\\
\begin{table}[htbp]
\begin{center}
\caption{Comparison of computation time for the beamlet-based and beamlet-free method averaged over 5 runs.}
\label{Tab:CompTime}
\begin{tabular}{c|c|c}
    \multicolumn{2}{c|}{Process} & $t$ [s] \\
    \hline
    \multirow{5}{*}{BL-based} & Plan Creation & $299\pm 2$\\
     & Beamlet Comp. & $3356\pm 24$\\
     & Optimization & $12601\pm 1714$\\
     & Final Dose Comp. & $301\pm 1$\\
     \cline{2-3}
     & \textbf{Total} & $\mathbf{16558\pm 1714}$\\
     \hline
    \multirow{3}{*}{BL-free} & Plan Creation & $299\pm 2$\\
     & Optimization & $4697\pm 129$\\
     \cline{2-3}
     & \textbf{Total} & $\mathbf{4996\pm 129}$\\
\end{tabular}
\end{center}
\end{table}
\section{Discussion}
A novel treatment planning method, not limited to proton therapy, has been introduced. It merges dose calculation and optimization into a single process, resulting in improved computational efficiency and accuracy. The so-called beamlet-free algorithm was able to achieve comparable plan quality to a conventional beamlet-based algorithm.\\
The comparability of the two methods is limited by the fact that different parameters have to be chosen for each of them. These parameters have to be chosen in a way that allows a balance between keeping computation time low enough to enable clinical workflows while achieving a high statistical confidence.  The statistical uncertainty inevitably present in MC-based treatment planning stems from the fact that MC methods, and in the case of the beamlet-free algorithm also the optimization, are based on stochastic processes. The two methods have different parameters that may be tuned to this end. For this comparison, the same number of primary protons was therefore used in the final dose calculation to ensure a comparable level of uncertainty for the MC simulation. The iteration number of the optimization algorithms is not comparable, since the direction of search is only partially estimated at each step of the beamlet-free algorithm which consequently needs a lot more iterations to converge. In both cases, the number of iterations utilized was considered as realistic to produce adequate plan quality in a reasonable time-frame. Another parameter that impacts computation time and quality of the results is the number of protons per beamlet in the beamlet-based method and the batch size in the beamlet-free. While they are hardly equivalent to one another, they play a similar role during the optimization. They are used as input for the optimization algorithm. Both need to be sufficiently high to prevent optimization based on inaccurate dose estimations, but still considerate of the total computation time. In beamlet-free treatment planning the three following parameters, total number of primary protons, number of iterations and batch size, depend on each other. However, they can be chosen independently in the beamlet-based case. This places some restrictions on the user, but might also be beneficial, since it reduces the risk of errors.\\
Both methods are able to fulfill all clinical objectives. Target coverage worsened slightly for the beamlet-free case, while OAR sparing improved for most organs. Altogether, the difference between the two methods is small, but statistically significant. It might stem from a number of sources. For example, a possible explanation could be misguided optimization steps based on insufficiently approximated gradients.\\
Regarding computational performances, the peak memory required during treatment planning could be reduced by $95\%$ through the use of the beamlet-free algorithm. The total computation time could be reduced by $70\%$ through use of the beamlet-free algorithm. Notice that the two methods profit unequally from the number of available computation threads. The MC code, MCsquare, utilizes CPU parallelization to improve computation time and is slowed down considerably, if the number of CPU threads is reduced. The beamlet-free algorithm also utilizes MCsquare for the dose simulation, but the optimization slows down the MC simulation itself. The beamlet-free algorithm still profits from having multiple CPU threads at its disposal, but especially very high numbers of threads cannot achieve comparable simulation times to the pure MC simulation. The update procedure of the mean target dose is here a bottleneck in the parallelization, that prevents the beamlet-free algorithm to draw the full benefit from very large numbers of CPU threads, probably due to conflicting accesses to the same chunks of memory. It should be noted that the presented data was obtained from simulations performed on a computation server with 80 CPU cores and that the benefit of the beamlet-free algorithm is considerably higher, for computations performed with a limited number of CPUs. For a more thorough comparison, both optimizers should be implemented within the same framework, to ensure identical conditions. Small differences can be expected by using an existing external code for the beamlet-based part (i.e. Scipy optimizer in our case) of the comparison.\\
The complexity analysis revealed that, in beamlet-based treatment planning, computation time and peak memory usage increase almost linearly with the number of spots. On the other hand, the beamlet-free method can achieve comparable plan quality without being affected by the number of spots, resulting in independent computation time and memory usage. The peak memory usage from the beamlet-free method was recorded below that of the beamlet-based simulation in all cases. With respect to computation time, for low numbers of spots the beamlet-based method outperformed the beamlet-free method, but for increasing numbers of spots, the beamlet-based method gets slowed down considerably. It is possible to determine roughly from which number of spots it would be beneficial to utilize the beamlet-free instead of a beamlet-based method. It should be kept in mind, however, that factors like the machine computations are performed on, the exact beamlet-based optimizer and implementation chosen for the comparison and the depth location of the target and number of simulated particles, can impact the two methods differently and render it impossible to fix any specific number.\\
It can be observed that for very low numbers of spots, the beamlet-free method can even bring disadvantages, but already for numbers of spots regularly encountered in clinical routines, the beamlet-free method starts to outperform conventional beamlet-based treatment planning. The benefits of the beamlet-free method are particularly highlighted in methods that require a large number of beamlets, such as robust optimization. In the robustness approach, traditional strategies make use of a set of worst-case scenarios \cite{Unkelbach_2007, Pflugfelder_2008}, compiling the treatments uncertainties. If a beamlet-based approach is used, the very costly influence dose matrix must be calculated for each scenario and this further brings a computational burden during the optimization. Additionally, a more recent development sees the emergence of arc proton therapy treatment modality \cite{DING20161107,gu2020novel,Engwall_2022,WUYCKENS2022105609} that delivers the dose across a very large set of irradiation angles, shaping the arc. However, increasing the number of beams translates into increasing the number of spots, resulting in producing a larger beamlet matrix than the ``classical'' one obtained with an IMPT treatment. Finally, the beamlet-free method could also play an interesting role for adaptive proton therapy where a new treatment plan might be needed in good time for an anatomy that has changed.\\

%In the future, we would like to incorporate robust optimization techniques into the %beamlet-free framework, as robustness has indeed become crucial in any proton treatment %plan given the large uncertainties on the proton range.\\

\section{Conclusion}
\label{sec:ccl}
The proposed beamlet-free algorithm could successfully generate treatment plans of comparable quality to conventional treatment planning methods. Memory requirements could be reduced by $95\%$ and the computation time by $70\%$ on an illustrative patient case. The scaling analysis revealed that the benefit to be drawn from the beamlet-free method increases for higher numbers of spots, which makes the method promising for new emerging treatment techniques such as robust optimization, adaptive and arc proton therapy.

\section{Aknowledgments}
D.P.~is funded through a Télévie grant by the F.R.S.-FNRS (Fonds de la Recherche scientifique). 
S.W.~is funded by the Walloon Region as part of the Arc Proton Therapy convention (Pôles Mecatech et Biowin). 
J.A.L.~is a Research Director with the F.R.S.-FNRS (Fonds de la Recherche scientifique).

\section{Conflict of Interest}
Kevin Souris is an employee at Ion Beam Applications SA.

%\section*{References}
%\begin{harvard}
%\end{harvard}
%\section*{References}
%\bibliographystyle{agsm}
%\bibliography{harvard}

\bibliographystyle{medphy}
%\bibliography{biblio}

\appendix
%\begin{figure*}[t]
%    \begin{center}
%        \includegraphics[height=8.5cm]{phantom.pdf}
%    \captionv{15}{}{Water phantom used for the scaling analysis. An examplary target region with side length of 20 mm is represented in red.
%    \label{fig:Phantom}}
%    \end{center}
%\end{figure*}
%\begin{figure*}[t]
%    \begin{center}
%        \includegraphics[height=8.5cm]{Time_contributions.pdf}
%    \captionv{15}{}{Time contributions of the individual treatment planning processes required for the beamlet-free method (left bar, striped) and the beamlet-based method (right bar, blank) for varying numbers of spots. 
%    \label{fig:Phantom}}
%    \end{center}
%\end{figure*}
%\begin{figure*}[t]
%    \begin{center}
%        \includegraphics[height=8.5cm]{TC.pdf}
%    \captionv{15}{}{Target coverage achieved by the beamlet-free method (blue) and the beamlet-based method (orange) for varying numbers of spots. 
%    \label{fig:Phantom}}
%    \end{center}
%\end{figure*}

\end{document}